 \documentclass[authoryear,preprint,review,12pt]{elsarticle}

\usepackage{amssymb}

\journal{Physics Letters}

\begin{document}

\begin{frontmatter}



\title{Electrons in Magnetic Mirror Geometry}


\author{R. A. Treumann$^{a,c}$ \& W. Baumjohann$^b$}

\address[1]{Department of Geophysics, Munich University, Munich, Germany}
\address[2]{Space Research Institute IWF, Austrian Academy of Sciences, Graz, Austria}
\address[3]{Visiting at International Space Science Institute, Bern}

\begin{abstract}
Landau's theory of electron motion in stationary magnetic fields is extended to the inclusion of bouncing along the field between mirror points in an inhomogeneous field. The problem can be treated perturbation theoretically.  As expected, bouncing is quantized, causes a weak shift in Landau levels, reduces parallel degeneracy, but does not contribute to diamagnetism.
\end{abstract}

\begin{keyword}
Magnetic bounce \sep Electrons in strong magnetic fields
\PACS 71.10.Di \sep 97.60.Gb


\end{keyword}

\end{frontmatter}


In a homogeneous magnetic field $\mathbf{B}=B_0\hat z$, with vector potential $\mathbf{A}(\mathbf{x})= (-B_0y, 0,0)$ in the Landau gauge, electrons perform Landau-cyclotron orbits around the magnetic field at cyclotron frequency $\omega_c=eB_0/m_e$.  The quantum mechanical problem had been treated long ago by Landau \citep{landau30} solving the Schr\"odinger equation \citep[see also][and others]{landau1965, kleinert2004} and yielding the celebrated quantization of perpendicular electron energy $\epsilon_\perp=\hbar\omega_c(L+\frac{1}{2})$ into the chain of Landau levels $L\in\textsf{N}$ and theory of electronic diamagnetism \citep[cf., e.g.,][]{landau30,Huang1965}. Clearly, in homogeneous magnetic fields, the parallel energy $\epsilon_\| =p_z^2/2m_e\equiv\hbar^2k_z^2/2m_e$ is unaffected. Since the problem is independent on $y, p_y$, Landau levels are highly degenerate. In a homogeneous field their degeneracy is $g= r_xr_z/\lambda_m^2$, where $r_x, r_z$ are the transverse to $y$ extensions of the spatial volume under consideration, i.e. the perpendicular surface expressed in terms of a magnetic length $\lambda_m= \sqrt{2\pi\hbar/eB_0}$ \citep{aharonov1959}. 

In a magnetic mirror geometry (like, e.g., the Earth's radiation belts) electrons of sufficiently low parallel energy  may become trapped, bouncing between the mirror points along the magnetic field \citep[cf., e.g.,][]{Kivelson1995,Baumjohann1996}. Classically this motion is understood as an oscillation along the magnetic field at bounce frequency $\omega _b$ \citep[cf., e.g.,][]{Hasegawa1975}. This frequency is formally given by the inverse double-bounce time $\tau_b$ which, for an electron of energy $\epsilon$, is an integral over the parallel energy $\epsilon_\|(s)=\epsilon-\epsilon_\perp(s)$ along the magnetic field:
\begin{equation}
\tau_b\equiv\frac{2\pi}{\omega_b}= \sqrt{\frac{2m_e}{\epsilon}}\int\limits_{s_1}^{s_2} \frac{\mathrm{d}s }{\sqrt{1-{\epsilon_\perp(s)}/{\epsilon}}}
\end{equation}
between the two mirror points $s_1,s_2$. These are defined as the points along the magnetic field, where all the electron energy is in the perpendicular, i.e. $\epsilon_\|(s_{1,2})=0$. For non-relativistic electrons, neglecting radiative losses, the electron conserves its magnetic moment $\mu=\epsilon_\perp(s)/B(s)$, and we have $\epsilon_\perp({s_{1,2}})=\epsilon$, or $\epsilon_\perp(s)/\epsilon=B(s)/B(s_{1,2})$. 

If we assume that the magnetic field possesses some approximate symmetry with respect to $s$ (for instance like a dipole field) and that the mirror points are located not too far away from the symmetry plane, then $s_1=-s_2\equiv -s_m$, and $B(0)\equiv B_0$ is the minimum of the magnetic field along the field. Expanding $B(s)$ around minimum, $B(z)\approx B_0(1+az^2)$. $a=\frac{1}{2}B''(s)|_{s=0}$ is the second derivative along $\mathbf{B}$ taken at field minimum. Close to the plane of symmetry $s\approx z$ becomes a straight coordinate $-z_m\leq z\leq z_m$ that is limited by $z_m=\sqrt{(R-1)/a}$, and $R=B(s_m)/B_0$ is the magnetic-field mirror ratio, the latter assumed being a known quantity.

With these approximations the vector potential of the mirror field becomes, in the Landau gauge, 
\begin{equation}\label{vecpot}
\mathbf{A}(x,z)=[-B_0(1+az^2)x,0,0], \qquad |z|\leq z_m.
\end{equation}
The correction introduced by the mirror geometry is small, and the vector potential still possesses only one component $A_y$ which, however, now depends on the two spatial coordinates $x,z$. This complicates the problem and, in addition, gives rise to a weak transverse magnetic component $B_y=\partial_zA_x=-2B_0ayz$. It causes a weak particle drift in $\mathbf{y}$-direction which, in higher order, would ultimately break the Landau degeneracy.

With the above vector potential the Hamiltonian of the gyrating and mirroring electron becomes
\begin{equation}\label{ham1}
{\cal H}= \frac{1}{2m_e}\left\{\Big[p_x-eB_0(1+az^2)y\Big]^2 +p_y^2+ p_z^2\right\}.
\end{equation}
Since, near field minimum $B_0$, $a$ and $z$ can be assumed small numbers, this expression can be substantially simplified by expanding the second term in the braces, yielding
\begin{equation}\label{ham2}
{\cal H}\approx {\cal H}_L + V(y,z).
\end{equation}
Here ${\cal H}_L= (2m_e)^{-1}\bigg\{p_y^2+p_z^2 +m_e^2\omega_c^2(y-y_0)^2\bigg\}$, with $y_0=p_x/m_e\omega_c$, is the usual harmonic oscillator Landau-Hamiltonian which (to lowest approximation) depends only on spatial coordinate $y$, and 
\begin{equation}\label{vpert}
V(y,z)= -\omega_caz^2y\Big(p_x-m_e\omega_cy\Big)
\end{equation}
is a perturbation which is considered to be small, a well justified assumption near field minimum along the field line. In the last two expessions $p_x, y, z$ are operators. Strictly speaking, $\omega_c(z)=eB_0\left(1+az^2\right)/m_e$ in ${\cal H}_L$ is a $z$-dependent cyclotron frequency. However, for the assumed small $a,z$ values, it varies only very weakly along the field. Including this $z$-dependence would account for a slight shift of the Landau levels and changes in the Landau wave functions along the field. In the following it will, for our limited purposes of investigating the more interesting effects of bounce motion, be taken as the constant value $\omega_c(0)=eB_0/m_e$ in the field minimum. 

Under these assumptions, the system is, by definition, periodic and thus stationary. With the Landau-Hamilton operator in the Schr\"odinger equation ${\cal H}_L |L(\mathbf{x})\rangle = \epsilon_L |L(\mathbf{x})\rangle$, the unperturbed Landau solution $|L(\mathbf{x})\rangle\equiv |L^{(0)}(\mathbf{x})\rangle$ is known \citep[see, e.g.,][and others]{landau30,landau1965,kittel1965}, and the whole problem can be treated by perturbation theory. We note that the perturbed Hamiltonian remains to be independent of $x$. Hence the corresponding momentum may be replaced by $p_x=\hbar k_x$, and degeneracy in $x$ remains unresolved even for a bouncing particle. Solving the perturbation problem will provide the energy shift 
$\Delta_L= \epsilon_L-\epsilon_L^{(0)}$ in the Landau levels; it also provides the perturbed wave function $|L^{(1)}\rangle \neq |L^{(0)}\rangle$.

Before proceeding, we instead determine the corresponding quantities for the pure bounce motion. This can be done by realizing that the Hamiltonian Eq. (\ref{ham1}) can be averaged over the fast Landau-cyclotron oscillations. This is easily done when writing classically for the two coordinates $\left\{ x(t), y(t)\right\}=r_c\left\{\cos\omega_ct,\sin\omega_ct\right\}$ with $r_c$ the classical gyroradius. Correspondingly, the momenta become $\left\{p_x,p_y\right\}=m_e\omega_cr_c\left\{-\sin\omega_ct,\cos\omega_ct\right\}$. Inserting into Eq. (\ref{ham1}) and averaging with respect to time over two gyroperiods $\tau_c=2\pi/\omega_c$ (because the Landau ground state frequency is $\frac{1}{2}\omega_c$) yields for the gyro-averaged pure bounce-Hamiltonian up to second order in $z$
\begin{equation}\label{hred}
{\cal H}_b= \frac{p_z^2}{2m_e}+\mu B_0(1+az^2),
\end{equation}
which is the Hamiltonian of a harmonic oscillator of reduced Hamiltonian ${\cal H}_\|={\cal H}_b-\mu B_0$ and energy eigenvalue $\epsilon'=\epsilon-\mu B_0$, with $\mu(L)=\epsilon_\perp(L)/B_0$ the magnetic moment of the electron in Landau level $L$. Its Schr\"odinger equation is ${\cal H}_\||\ell\rangle= (\ell+\frac{1}{2})\hbar \omega_\||\ell\rangle$, which identifies 
\begin{equation}
\epsilon'_\ell(L)=\left(\ell+\textstyle{\frac{1}{2}}\right)\hbar\omega_\|(L)
\end{equation}
as the energy spectrum of the bounce oscillation in Landau state $L$, and 
\begin{equation}
\omega_\|(L)=\sqrt{\frac{a\mu(L) B_0}{m_e}}=\sqrt{\frac{a\hbar\omega_c}{m_e}\left(L+{\textstyle\frac{1}{2}}\right)}\ll\omega_c 
\end{equation}
as the corresponding quantum mechanical bounce frequency. The wave function $|\ell\rangle$ can be found in any QM textbook \citep[see, for instance,][]{Sakurai1994}. As expected, the parallel energy of an electron in any Landau state $L$ bouncing along the magnetic field is quantized. The restriction on the parallel quantum number is $\ell/L<\omega_c/\omega_\|$ imposing an upper bound on $\ell$ for every Landau quantum number $L$. Since $\omega_c\gg\omega_\|$, any perpendicular Landau level contains many parallel bounce levels. This resolves the parallel degeneracy reducing Landau degeneracy to $g=r_x/\lambda_m$. Though this is not a particularly deep insight, it clarifies the properties of electron motion in mirror magnetic field geometries. 

One may also ask whether the bounce motion in an electron gas of density $N$ introduces any contribution $\chi_b$ to the diamagnetic susceptibility $\chi$. This correction is obtained from the second derivative of the logarithm of the bounce partition function  $\log{\cal Z}_b$ with respect to the magnetic field $B$ at constant temperature $T$, volume $V$, density $N$, and chemical potential $\bar\mu$. It can be shown by conventional methods \citep[cf., e.g.,][]{Huang1965} that
\begin{equation}
\log{\cal Z}_b\approx NV\left[1-\frac{1}{24}\left(\frac{\hbar\omega_\|}{T}\right)^2\right].
\end{equation}
Inserting for $\omega_\|$ one immediately finds $\left(\partial^2\log{\cal Z}_b/\partial B_0^2\right)|_{T,V,\bar\mu}=0$, hence yielding $\chi_b=0$. To lowest order the bounce motion does not contribute to magnetic susceptibility. This is reasonable, because contrary to Landau orbits bouncing electrons do not cause a net current, whose magnetic effect thus vanishes.

Returning to our initial problem, we may now determine the perturbation of any of the Landau levels caused by the parallel electron motion when the electron is forced to perform a bounce along a magnetic mirror field geometry.  We are mainly interested in the energy shift  
\begin{equation}\label{energyshift}
\Delta_L^{(1)} = \langle L^{(0)}|V|L^{(0)}\rangle \approx \frac {1}{3}\frac{\hbar^2 k_x^2}{m_e}az_m^2 \Big[f\!\left(L\right)\Big].
\end{equation}
of a particular Landau level caused by the bounce motion. This can, to lowest perturbation order, be calculated simply from the known unperturbed Landau wave function $|L^{(0)}\rangle$ \citep[see, e.g.,][the latter for its path-integral amplitude]{landau1965,kittel1965,kleinert2004}Ê\ using the perturbation Eq. (\ref{vpert}) of the Hamiltonian. The Landau wave function consists of products of exponentials and Hermite polynomials. Exploiting the orthonormality conditions in $y$, the term proportional to $yp_x$ in $V(y,z)$ is linear in $y$; being odd it vanishes by $y$-integration. The contribution comes from the quadratic in $z$ and $y$ term $+m_e\omega_c^2az^2y^2$. Dimensionless $z$-integration generates a factor $\frac{1}{3}m_eaz_m^2\omega_c^2$. Integration with respect to $y$ adds the factor $\big[f(L)\big]$ in Eq. (\ref{energyshift}), where
\begin{equation}
\Big[f\!\left(L\right)\Big]=1+ \frac{1}{\lambda_m^2k_x^2}\left[1+\frac{1}{2}\frac{(L-2)!}{L!}\left(\frac{5L}{2}-\frac{L+2}{2^L}-1\right)\right]
\end{equation}
is obtained by conventional methods \citep{gradsteyn65} and after some simple algebra. Here $\lambda_m=\sqrt{\hbar/eB_0}$ is the `magnetic length' \citep{aharonov1959}. The $L$-dependence in the brackets vanishes for Landau levels with $L\leq 2$. The two lowest order Landau levels experience a constant energy shift. The final result for the Landau energy shift is the approximate expression given on the right in Eq. (\ref{energyshift}). Since $a>0$ (being the second derivative of $B(s)$ in the minimum) is always positive, the shift in perpendicular energy caused by the bounce perturbation on a Landau level is positive and small. This follows from the rapid decay with increasing $L$ of the negative term in parentheses.  In general, the smallness of the correction is due to the pre-factor in the second term in $f(L)$; the correction decays the higher the Landau level.

To first and higher orders the bounce also introduces a weak $z$-dependence of the perturbed Landau wave function which destroys the parallel degeneracy. Its exact expression, taking care of the degeneracy, can be calculated by conventional perturbation methods, being here, however, of lesser interest as long as no practical application is urgent. In such a case (as for instance in the case of a very strong magnetic dipole field like in pulsars or magnetars), one would rather include the correct functional $s$-dependence of the magnetic field. Here we have just shown the expected dominant effect in the lowest order energy shift when electrons bounce in a magnetic mirror field geometry.

\subsection*{Acknowledgement}
{Hospitality of the ISSI staff during a brief visiting period of  RT is thankfully acknowledged.}




\begin{thebibliography}{99}

\bibitem[Aharonov \& Bohm(1959)]{aharonov1959}
Aharonov A. \& Bohm D., \emph{Phys. Rev.} \textbf{115}, 485-491 (1959) doi:10.1103/PhysRev.115.485.






\bibitem[Baumjohann \& Treumann(1996)]{Baumjohann1996}
Baumjohann W. \& Treumann R. A., \emph{Basic Space Plasma Physics} (Imperial College Press, London 1996) ch. 2.5.

\bibitem[Gradshteyn \& Ryzhik(1965)]{gradsteyn65}
Gradshteyn I. S. \& Ryzhik I. M., \emph{Table of Integrals, Seires, and Products}, 4th Edition (Academic Press, New York 1965) ch. 7.37.


\bibitem[Hasegawa(1975)]{Hasegawa1975}
Hasegawa A., \emph{Plasma Instabilities and Nonlinear Effects}, (Springer Verlag, Berlin-Heidelberg-New York, 1975) ch. 2.2.


\bibitem[Huang(1987)]{Huang1965}
Huang K., \emph{Statistical Mechanics}, 2nd Edition (John Wiley \& Sons, New York, 1987) ch. 11.

\bibitem[Kivelson \& Russell(1995)]{Kivelson1995}
Kivelson M. G. \& Russell C. T., \emph{Introduction to Space Physics}  (Cambridge University Press, Cambridge UK, 1995) ch.10.


\bibitem[Kittel(1963)]{kittel1965}
Kittel C., \emph{Quantum Theory of Solids} (John Wiley \& Sons, New York, 1963) ch. 11.


\bibitem[Kleinert(2004)]{kleinert2004}
Kleinert H., \emph{Path Integrals in Quantum Mechanics, Statistics, Polymer Physics, and Financial Markets}, 3rd edition (World Scientific Publ. Comp., Singapore 2004) ch. 9.


\bibitem[Landau(1930)]{landau30}
Landau L. D., Zeitschrift f\"ur Physik \textbf{64}, 629-637 (1930).

\bibitem[Landau \& Lifschitz(1965)]{landau1965}
Landau L. D. \& Lifshitz E. M., \emph{Quantum Mechanics} (Pergamon Press, New York, 1965) ch. 15.


\bibitem[Sakurai(1994)]{Sakurai1994}
Sakurai J. J., \emph{Modern Quantum Mechanics}, Revised Edition (Addison-Wesley Publ. Comp., Reading, Mass. 1994) chs. 2 and 5.


\end{thebibliography}
\end{document}